\documentclass{vienna-conf2019}
\usepackage{graphicx}
\usepackage{hyperref}
\usepackage[]{natbib}  
\usepackage{epstopdf}

\def\BibTeX{{\rm B\kern-.05em{\sc i\kern-.025em b}\kern-.08em
    T\kern-.1667em\lower.7ex\hbox{E}\kern-.125emX}}
\bibpunct{(}{)}{;}{a}{}{,}  

\begin{document}

\TitreGlobal{Stars and their variability observed from space}


\title{Challenges to modelling from groundbreaking new data of
present/future space and ground facilities}

\runningtitle{Challenges from new data}

\author{G. Clementini}\address{Osservatorio di Astrofisica e Scienza dello Spazio, Via Piero Gobetti 93/3, 40129 Bologna, Italy}

\author{M. Marconi}\address{Osservatorio Astronomico di Capodimonte, Via Moiariello 16, 80131, Naples, Italy}

\author{A. Garofalo$^1$}




\setcounter{page}{237}


\maketitle
\begin{abstract}
The sheer size of high-accuracy, multi-band photometry, spectroscopy, astrometry and seismic data that 
space missions like Kepler, Gaia,  
PLATO, TESS, JWST and ground-based facilities under development  such as MOONS, WEAVE and the LSST will produce within the next decade,  brings big opportunities to improve current modelling; but it also presents unprecedented challenges to overcome the present limitations in  
stellar evolution and pulsation models.
Such an unprecedented harvest of data also requires multi-tasking  and synergic approaches to be interpreted  and fully exploited. 
We briefly review major outputs expected from ongoing/planned facilities and large sky surveys, then focus specifically on Gaia and present a few examples of the  impact that this mission is having on studies of stellar physics, Galactic structure and the cosmic distance ladder.

\end{abstract}



\section{Introduction}
The wealth and  variety of  datasets 
produced by  ground/space-based facilities and large sky surveys under way or planned for the near future make  Astronomy a paradigmatic example of ``Big Data" science.  Some of these facilities are briefly reviewed,  
showing 
how their complementary data products can not only significantly  advance our knowledge of the stellar interiors, stellar evolution and pulsation, but can also help constraining the structure and formation of our Galaxy, can allow us to  characterise the stellar populations in Galactic and extragalactic environments and  set the cosmic distance ladder and the gauge of the expansion rate of the Universe. 

Past, ongoing and future space facilities  like WIRE \citep{Hacking99}, MOST \citep{Walker03}, CoRoT \citep{Baglin03}, Kepler \citep{Koch10}, K2 \citep{Howell14}, BRITE \citep{Pablo16}, TESS \citep{Ricker15}, Cheops \citep{Broeg13}, and PLATO \citep{Rauer14} are producing unprecedented, accurate light curves, which reveal a very rich spectra of stellar oscillations (gravity-modes, pressure-modes, rotation-related modes, etc.) for stars in different evolutionary stages. Asteroseismology is exploiting these data to provide  seismic measurements of radii, masses, ages, distances and position on the Hertzsprung-Russell diagram for thousands of stars within some tens of kpc. 
These measurements allows us to test  distances from 
parallaxes, such as those measured by Gaia, and also provide a unique benchmark for testing and improving stellar evolution and pulsation models.          
On the theoretical side, 3D atmosphere models now start to become available. They will enable a calibration of the empirical oscillations and the convection parameters used in stellar evolution codes
\citep[see e.g.][and various contributions in this conference]{Chaplin13,Dupret19}.

Several physical mechanisms (e.g. rotation and rotationally-induced mixings,  magnetic fields, thermohaline mixings, internal gravity waves, mass loss, etc.) are still poorly known  and are  not properly accounted for in current stellar modelling. 
 Inclusion of such effects into models is difficult because they are controlled by several physical parameters. However, it is no longer possible to ignore them if we wish to properly understand and interpret the huge amount of high-accuracy photometric, spectroscopic, astrometric and seismic  information that on-going and future surveys are providing. The development of 3D stellar  models  and hydro-dynamical codes is  also needed to realistically describe convection and other dynamical phenomena occurring in stars, however, this is a very challenging and computationally expensive task \citep[see e.g.][and references therein, and a number of talks at this conference]{Joyce19}.

Large time-domain photometric surveys such as OGLE \citep{Udalski92}, MACHO \citep{Alcock99}, EROS \citep{Tisserand07}, ASAS \citep{Pojmanski97}, Pan-STARRS \citep{Chambers16}, Hipparcos \citep{vanLeeuwen07}, SDSS (Stripe82; \citealt{Annis14}), Catalina \citep{Drake14}, PTF \citep{Law09}, ZTF \citep{Bellm19}, VVV \citep{Minniti10}, VMC \citep{Cioni11} and Gaia \citep{Prusti16} are providing a census of the variable stars in the Milky Way and its closest companions, un-disclosing new features and new variability types. 
Starting full science operations in 2023, LSST (\citealt{Ivezic19} and references therein) will be Gaia's deep complement in the south hemisphere,  providing parallaxes, proper-motions, and multiband photometry with similar uncertainties than at Gaia's faint end (V$\sim$20.5 mag) but  up to about five magnitudes fainter than Gaia. 

In parallel, large spectroscopic surveys such as 
SEGUE \citep{Yanny09}, RAVE \citep{Steinmetz06}, GALAH \citep{Desilva15}, APOGEE \citep{AllendePrieto08}, LAMOST \citep{Deng12} are measuring radial velocities and elemental abundances. 
Gaia-ESO \citep{Gilmore12},  the only spectroscopic survey at an 8 m class telescope so far,  is providing large samples of heavy elements abundances, both for s-process-dominated (Y-Zr-Ba-La-Ce), r-process-dominated (Sm-Eu), and mixed s\&r (Pr-Nd) elements. In the near future, instruments under development such as WEAVE \citep{Bonifacio16} at the William Herschel Telescope (WHT), 4MOST \citep{deJong19} at VISTA and MOONS \citep{Cirasuolo11} at the VLT will provide a detailed chemical characterisation for millions  of stars in the Galactic halo and disk(s) (WEAVE) 
and  accurate  chemistry and kinematics for large samples of old giants spanning a wide portion of the red giant branch in the Magellanic Clouds and the Sagittarius  dwarf spheroidal (MOONS). 
Accurate nucleosynthesis predictions,  in particular for s-elements,  require a detailed  modelling of the asymptotic giant branch (AGB) evolutionary phase. 
The AGB phase is also critical for the interpretation of the infrared observations of the evolved populations in galaxies (e.g. the Magellanic Clouds and Local Group dwarf galaxies) and to study  extinction properties.          
In the future, study of the circumstellar envelopes around massive stars in the Milky Way and in the Local Group will also become possible thanks to the high spatial resolution of next generation facilities such as ELT in the  optical and SKA in the radio. SKA will also allow to measure the magnetic field in  stars of different evolutionary phases.

 \begin{figure}[ht!]
 \centering
  \includegraphics[width=0.48\textwidth,clip]{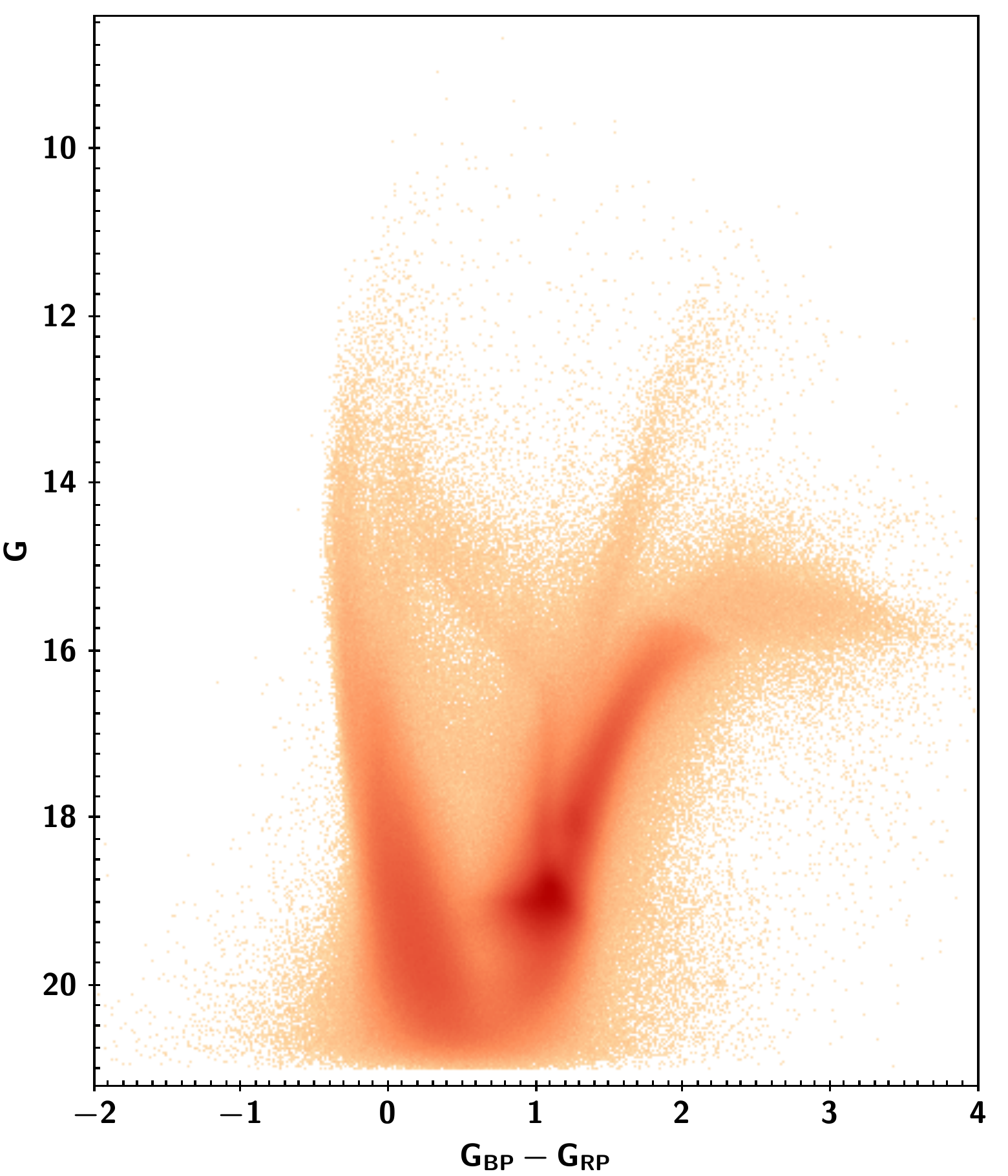}
  \includegraphics[width=0.48\textwidth,clip]{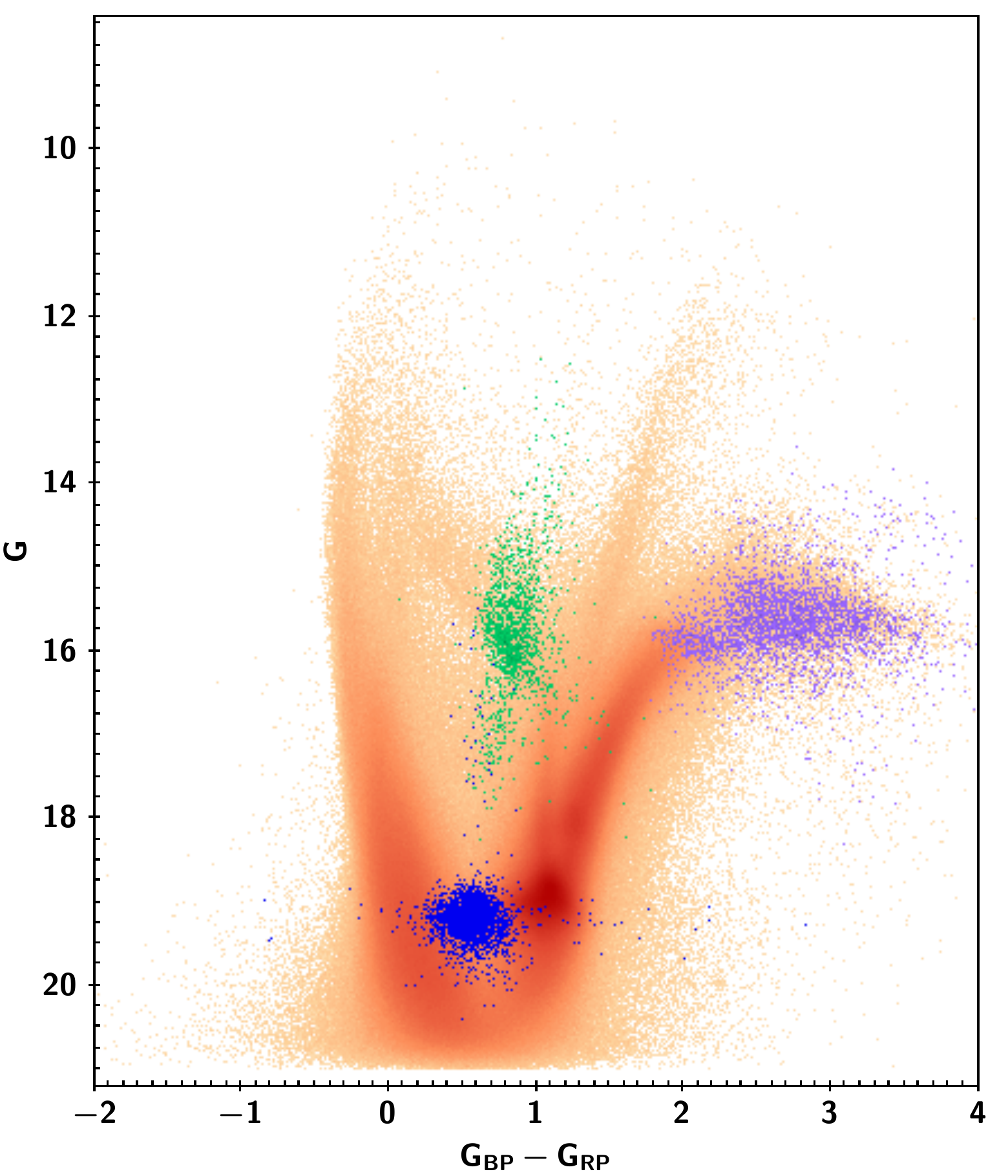}    \caption{{\bf Left:} CMD of sources in the Gaia DR2 catalogue (selected by parallax and proper motions)  contained in a region of about 8.8 degrees in radius around the Large Magellanic Cloud centre. Main features corresponding to the different evolutionary phases can be easily recognised: the main sequence, the red giant branch, the red clump, the horizontal branch. A sharp cut on the red giant branch marks the tip of the red giant branch (TRGB). The magnitude of the TRGB, which is set by the luminosity of the He core flash, may serve as a distance indicator for old stellar populations.  Also very clearly visible are  the blue loops   of the core helium-burning evolutionary phase.
  {\bf Right:} Same as in the right panel but with Cepheids, RR Lyrae stars, and LPVs \citep[respectively]{Soszynski17,Soszynski16,Soszynski09} that cross-match with variable sources in the DR2 catalogue plotted as green, blue and purple filled circles, respectively, using G, G$_{BP}-$G$_{RP}$ mean magnitudes and colours from Gaia DR2 variability tables. The bulk of classical Cepheids places on the central helium burning blue loop evolutionary phase, whereas RRLs nicely trace the LMC horizontal branch.  Most of the LPVs are found above the TRGB, in the region of thermally-pulsing AGB stars.}
 \label{clementini:fig1}
\end{figure}

Such an unprecedented harvest of data requires synergic and multivariate approaches to be  fully exploited.

\section{Gaia, three instruments in one mission: astrometry, (spectro-)photometry, spectroscopy} 
The stunning revolution being operated by Gaia has often been mentioned during the conference and examples have been shown in a number of talks (see, e.g. Eyer et al. contribution to this proceedings). Here, we would like to address two specific fields where Gaia is really astonishing: (i) the detailed monitoring of stellar populations in different evolutionary phases, and (ii) the distance scale.

The study of stellar populations can rely on Gaia 3-band time-series photometry
(G, G$_{BP}$ and G$_{RP}$) and G$_{BP}$, G$_{RP}$ spectro-photometry; on spectroscopy from the Radial Velocity Spectoctrometer (RVS; for sources brighter than G$\sim$ 16-16.5 mag) and on astrometry (positions, proper motions and parallaxes, hence individual distances) for over 1 billion stars, that allow us to build accurate  HR diagrams (see left panel of Fig.~\ref{clementini:fig1}, showing the  
colour magnitude diagram of the Large Magellanic Cloud -- LMC, from Gaia Data Release 2 -- DR2, data) as well as to estimate precise individual and mean distances. On this basis a 3D map of the Milky Way providing insight into the Galactic formation and evolution mechanisms is derived. Gaia is also a most powerful tool to discover and characterise all-sky variable sources, as  shown by the catalogue and multiband time-series for more than  half a million variables of different types (RR Lyrae stars, Cepheids,  Long Period Variables -- LPVs, Solar-like stars with rotation modulation, $\delta$ Scuti \& SX Phoenicis and short period variables) released in Gaia DR2 \citep{Holl18}. In the right panel of Fig.~\ref{clementini:fig1}, we show RR Lyrae stars, Cepheids and LPVs released in DR2 which belong to the LMC plotted over the galaxy CMD. 

A noteworthy product released in DR2, is the catalogue of about 150,000 rotational modulation variable candidates of the BY Draconis class, an  unprecedented sample to study star rotation,  magnetic activity and stellar ages \citep{Lanzafame18,Lanzafame19}.

Gaia DR2 contains also a catalogue of 140,784 confirmed  RR Lyrae stars with full characterization: periods, amplitudes, mean magnitudes, Fourier parameters, photometric metal abundances (for a subsample of 64,932 sources)  and interstellar absorption (for 54,272 of them) \citep{Clementini19}. About 50,000 of these RR Lyrae stars  are new discoveries by Gaia. In the DR2 variability tables there are more than 9,000  confirmed Cepheids fully characterized and with photometric metal abundances (for 3,738 for sources). About 350 of them are new discoveries by Gaia \citep{Clementini19, Ripepi19}.

\begin{figure}[ht!]
 \centering
 \includegraphics[width=0.7\textwidth,clip]{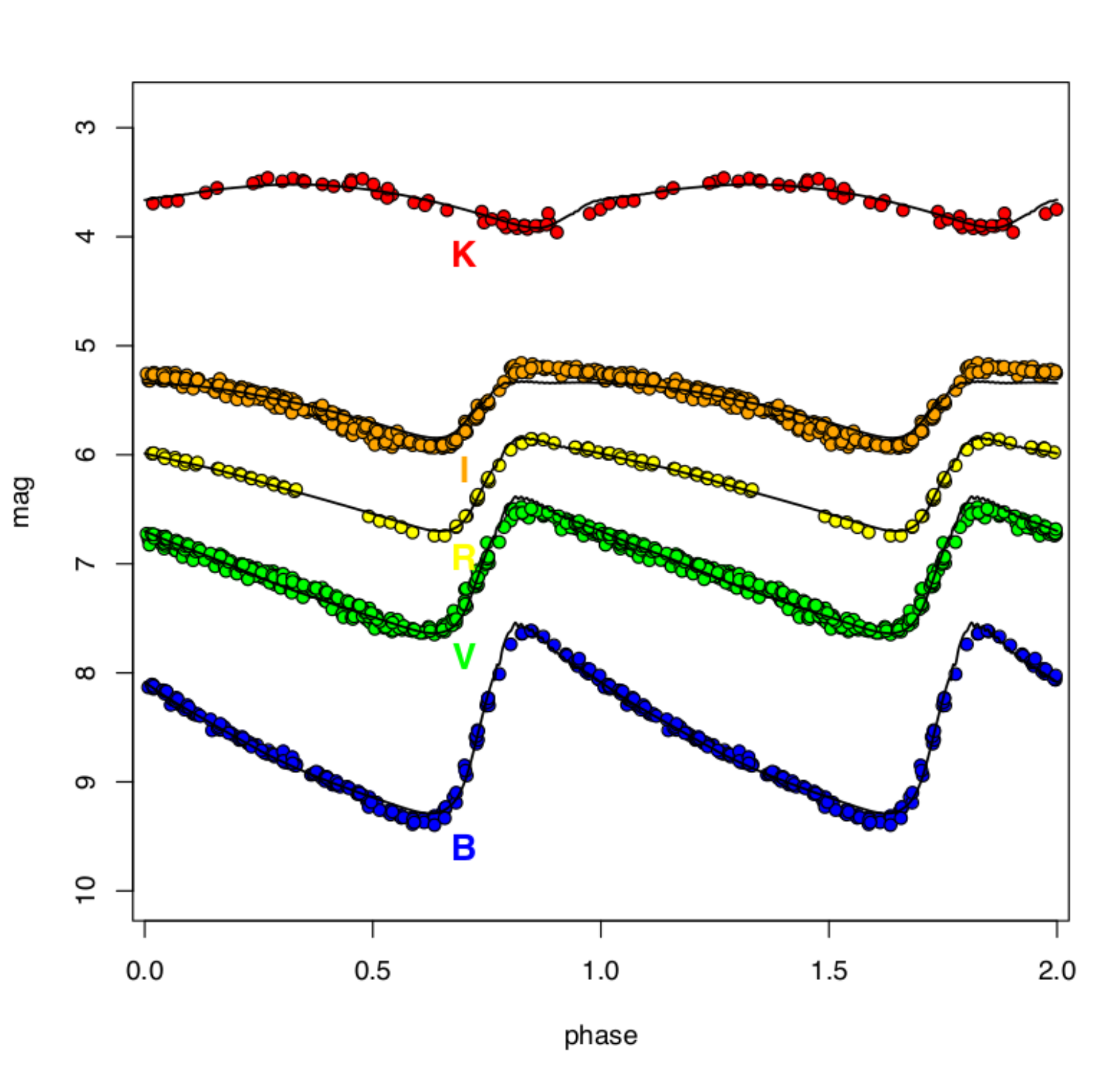}      
  \caption{Model fitting of the multi-wavelength light curves of the fundamental mode classical Cepheid RS Puppis (P=41.528 days) through non-linear convective pulsation models 
  (adapted from Fig.~12 of \citealt{GaiaColl17}).}
  \label{clementini:fig2}
\end{figure}

In the case of pulsating stars, the distance information inferred from Gaia parallaxes can be used to provide stringent constraints on other debated quantities and relations such as the efficiency of superadiabatic convection or Mass-Luminosity relations, as well as, once complementary spectroscopic metallicities are available, the Helium  to metal enrichment ratio. This will be possible through the comparison between observed and predicted pulsation properties including the model fitting of multi-filter light curves through non-linear convective pulsation models \citep[see e.g.][and references therein]{Marconi05,Keller06,Marconi13a,Marconi2013b}. 
Figure~\ref{clementini:fig2}  
 shows the model fitting of the multiband light curves of the classical Cepheid RS Puppis. The best fit is obtained for a model with  T$_{\rm eff}$ = 4875 K, $\log$ L/L$_\odot$ = 4.19, M/M$_{\odot}$=9 and mixing-lenght parameter $\alpha$ = 1.5. The model fitting 
  provides a parallax of 0.58$\pm$ 0.03 mas in excellent agreement with the Tycho-Gaia Astrometric Solution (TGAS)  parallax released with Gaia DR1 for this star, 0.63$\pm$ 0.26 mas. 

Furthermore, once fixed the distances, it will be possible to constrain the coefficients of the adopted extinction laws in current applications of the Period-Wesenheit relations.
The comparison between predicted and observed radial velocity curves or Period-Radius relations will also allow us to directly measure the projection factor  P, whose value and possible dependence on the pulsation period are debated in the literature.
 

Gaia distances are also adopted to calibrate Cepheid Period-Luminosity and Period-Wesenheit relations that in turn allow us to calibrate the extra-galactic distance ladder and to evaluate the Hubble constant 
 through the calibration of secondary distance indicators.

\section{Parallaxes - Distances - H$_0$ and the H$_0$ `tension'}


\begin{figure}[ht!]
 \centering
 \includegraphics[width=0.9\textwidth,clip]{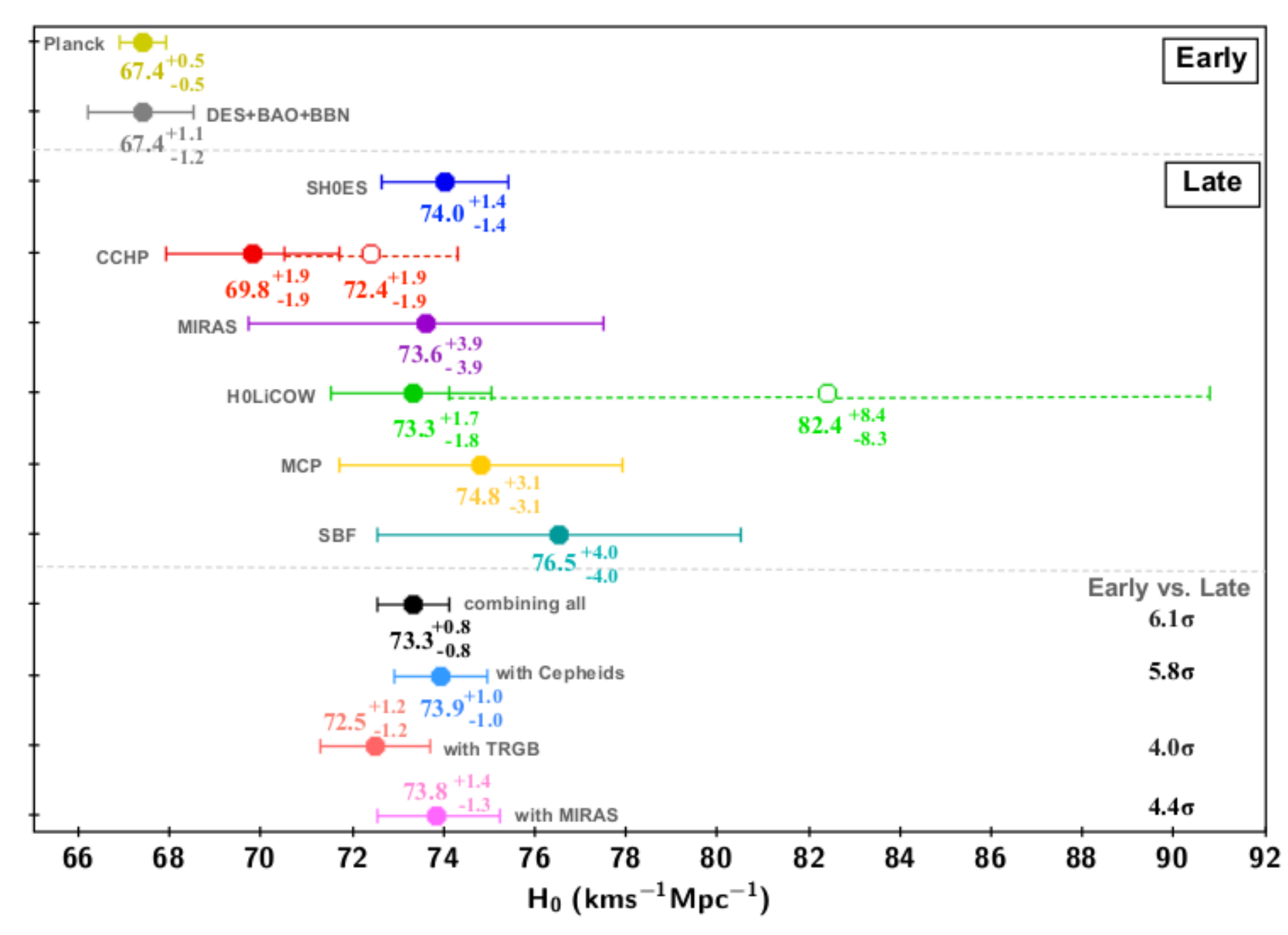}      
  \caption{Predictions and measurements of H$_{0}$ based on early- and late-Universe probes (adapted from Fig. 1 in \citealt{Verde19}). The two independent predictions based on early-Universe data are from: \citet{PlanckCollaboration18}, and \citet{Abbott18}, respectively. Results based on late-Universe data include: \citet{Riess19} results for the SH0ES collaboration which uses geometric distances to calibrate Cepheids; \citet{Freedman19} results for the CCHP collaboration which uses the TRGB to connect the distance ladder and,  shown by the red dashed line,  
\citet{Jee19} revision of the CCHP TRGB-based estimate of H$_{0}$;  \citet{Huang18} and \citet{Huang19} results for Miras in NGC 4258 and NGC 1159, respectively; \citet{Wong19} for the H0liCOW team that uses strong lensing time delays between multiple images of background quasars and, shown  by the green dashed line,  
the new measurements from strong gravitational lenses  by \citet{Jee19} who recently increased the H$_{0}$ value by H0liCOW to H$_{0}$=82.4 +8.4/-8.3 km s$^{-1}$ Mpc$^{-1}$ (but note the large uncertainty); 
new results from the Megamaser Cosmology Project (MCP; \citealt{Reid09}) which uses VLBI observations of water masers orbiting around supermassive black holes to measure geometric distances; and, \citet{Potter18} results from IR Surface Brightness Fluctuations. Not shown in the figure, 
but potentially an additional tool to measure H$_{0}$, are gravitational waves and standard sirens. Recent results from this method were published by \citet{Mukherjee19} who find H$_{0}$ = 69.3$^{+4.5} _{-4.0}$ km s$^{-1}$ Mpc$^{-1}$.}
  \label{clementini:fig3}
\end{figure}

The Hubble constant, H$_{0}$, is the expansion rate of the Universe measured in units of inverse time. There is `tension' between values of H$_{0}$ as derived from measurements of the anisotropy of the cosmic microwave background (CMB) radiation and as measured from a series of distance indicators in the local Universe. 
The CMB measures the age of the Universe at recombination. The distance ladder measures the age of the Universe now. If the standard model of cosmology is correct, these measurements should agree on the value of H$_{0}$ within the errors.
Figure~\ref{clementini:fig3} summarises values of H$_{0}$ based on early- and late-Universe probes presented in the conference: ``Tensions between the Early and the Late Universe", held at the Kavli Institute for Theoretical Physics, UC Santa Barbara on July 2019. 
The figure is an updated version of Fig. 1 in \citet{Verde19} and shows that currently that `tension' is between 4$\sigma$ and 5.8$\sigma$. 
One thing is clear from Fig.~\ref{clementini:fig3} the onus to improve the accuracy of the  H$_0$ measurements is on the distance ladder, rather than on the CMB. The error budget of the distance ladder must therefore be fully understood. 


Gaia contribution to understanding and quantifying the H$_{0}$ tension, as arising from the distance ladder side, will be unprecedented. This  mission  will allow us  to raise the accuracy of the astronomical distance ladder  by specifically tackling uncertainties and systematics in main stellar standard candles in order to cast light on the origin of the tension and at the same time better understand the underlying stellar physics.

The accuracy of local H$_{0}$ determinations will be significantly improved already by building on the data products in the forthcoming Gaia Data Releases 3 (EDR3 in the second half of 2020 and DR3 in the second half of 2021), and further boosted by subsequent releases (Gaia DR4, likely to occur in 2024) and the combination with data from TESS, JWST \citep{Beichman12} and the LSST. 

Gaia will specifically improve the `Anchors' of the distance ladder by directly measuring their distances through parallaxes. 
Examples of these improvements were already shown by the TGAS parallaxes released in Gaia DR1 (see, e.g. \citealt{GaiaColl17}) and by the Gaia-only parallaxes of RR Lyrae stars \citep{Muraveva18} and Cepheids \citep{Riess18} released in Gaia DR2.



According to current estimates of the error budget associated to each step of the cosmic ladder,  the improvement that Gaia is going to provide can allow us to evaluate H$_{0}$   to $\sim$ 1\%.
This will occur through a number of progressive steps that are briefly listed below:
\begin{itemize}
\item The exploitation of Gaia DR3 parallaxes along with a detailed investigation of the associated systematics, offsets (e.g. the offset with respect to QSO, see fig. 12 in \citealt{Lindegren18}) and relativistic effects, also relying on the comparison with asteroseismic parallaxes from Kepler, K2 and TESS
\item The use of Gaia parallaxes along with NIR photometry for pulsating stars  to re-calibrate Cepheid and RR Lyrae distance scales and their application to measure the distance to stellar systems containing different, independent primary and secondary distance indicators and,  at the same time, bridging Gaia's distance range to the LSST and JWST ones in a self-consistent path to H$_{0}$
\item The simultaneous development and extension of fine grids of nonlinear convective pulsation models for variable stars in different evolutionary phases and environments that will allow us to theoretically constrain the distances and their dependence on physical and numerical assumptions, with relevant implications for the final error budget associated to H$_{0}$
\item An improved treatment of population effects in  various classes of standard candles associated to different stellar populations, namely, Cepheids, RR Lyrae stars, LPVs and the TRGB, directly calibrated through Gaia parallaxes locally, and all well represented in many external systems, like the Magellanic Clouds 
\item The investigation of possible alternative cosmic distance scale anchors to the traditionally adopted LMC, such as M31, and the use of different independent indicators for the same stellar system/anchor
\item  A rigorous quantification of systematic effects associated to the various adopted distance indicators and their impact on the final H$_{0}$ derivation.
\end{itemize}

\section{Conclusions}
Synergy between different techniques, instruments and datasets  is the key to tackle
many of the issues  affecting stellar  evolution and pulsation  modelling as well as to test  empirical results (e.g. Gaia parallaxes).
A bright future is in front of us thanks to present/future outstanding facilities and surveys, providing an unprecedented wealth of excellent photometriy/astrometry/spectroscopy/asterosysmology datasets to challenge stellar evolution and pulsation modelling.

\begin{acknowledgements}
Acknowledgement: This work has made use of data from the European Space Agency (ESA) mission
{\it Gaia} (\url{https://www.cosmos.esa.int/gaia}), processed by the {\it Gaia}
Data Processing and Analysis Consortium (DPAC,
\url{https://www.cosmos.esa.int/web/gaia/dpac/consortium}). Funding for the DPAC
has been provided by national institutions, in particular the institutions
participating in the {\it Gaia} Multilateral Agreement. We wish to warmly thank our colleagues and collaborators in the project ``SH$_0$T: the Stellar path to the H$_0$ Tension” with whom many of the 
ideas presented in this proceedings are being developed. \end{acknowledgements}

\bibliographystyle{aa}  
\bibliography{clementini_5k04} 

\end{document}